# Investigation on high-order planar Hall effect in trigonal PtBi$_2$

Fangqi Cai,[1] Mingxi Chi,[2] Yingjie Hu,[1] Heyao Liu,[2] Yangyang Chen,[1,3] Chao Jing,[1,a)] Wei Ren,[1,a)] and He Wang[2,a)]

[1]Department of Physics, Materials Genome Institute, Shanghai Key Laboratory of High Temperature Superconductors, International Centre of Quantum and Molecular Structures, Shanghai University, Shanghai 200444, China

[2]Center for Quantum Physics and Intelligent Science, Department of Physics, Capital Normal University, Beijing 100048, China

[3]2-Dimensional Crystal Consortium, Pennsylvania State University, University Park, Pennsylvania 16802, USA

[a)]**Author to whom correspondence should be addressed:** cjing@staff.shu.edu.cn; renwei@shu.edu.cn; and wanghe@cnu.edu.cn.

# ABSTRACT

The trigonal $PtBi_2$ (t-$PtBi_2$) as a Weyl semimetal possessing triply degenerate points in its electronic bands near the Fermi level endows it with rich electronic properties. Previous studies have already measured the planar Hall effect (PHE) and in-plane anisotropic magnetoresistance (AMR) of t-$PtBi_2$. We noticed that their experimental results exhibited high-order features in both the PHE and AMR, yet these features were not systematically investigated. In our work, we conducted more systematic measurements and analyses of the PHE and AMR in t-$PtBi_2$. Both PHE and AMR show high-order features under low temperatures and strong magnetic fields, and these features share a similar temperature and magnetic field dependence with the turn-on behavior of resistance and temperature curves, indicating a common physical origin for them. We further summarize the critical conditions for the emergence of high-order PHE in t-$PtBi_2$, which will help to understand the origin of high-order features. In addition, we performed computational simulations on the AMR of t-$PtBi_2$, and the results were consistent with the experiments, indicating the high-order features are the result of the combined contribution of the Fermi surface anisotropy and the scaling behavior of magnetoresistance. Our findings will contribute to a deeper understanding of the origins of high-order features in non-magnetic topological materials.

Recently, trigonal PtBi$_2$ (t-PtBi$_2$), a non-magnetic topological semimetal with a layered structure, has garnered significant attention in condensed matter physics.[1–14] Its unique electronic structure, featuring Weyl points and triple degenerate points,[2] endows t-PtBi$_2$ with a plethora of electronic properties, such as extremely large magnetoresistance,[2] strong Rashba-like spin splitting,[3] topological edge states,[6] and negative magnetoresistance (NMR) induced by chiral anomaly.[10] Additionally, t-PtBi$_2$ is widely regarded as a promising candidate for realizing topological superconductivity,[5,9,12–14] just like other topological (semi)metals.[15–17] t-PtBi$_2$ is a kind of type I Weyl semimetal, which is predicted to exhibit the NMR and planar Hall effect (PHE) due to the chiral anomaly arising from nontrivial Berry curvature in momentum space.[18,19] The PHE, a specific transport phenomenon occurring when a magnetic field is coplanar with the Hall electric field, has been extensively observed in various topological materials.[20–27] Notably, a significant PHE was recently reported in t-PtBi$_2$ single crystals,[10] primarily attributed to the chiral anomaly and in-plane orbit magnetoresistance (OMR)[28,29] after ruling out other potential mechanisms such as current jetting,[30] and anisotropic magnetic scattering.[31] We further noticed that in previous PHE measurements of t-PtBi$_2$, both the PHE and anisotropic magnetoresistance (AMR) curves exhibited high-order features under low-temperature and high-magnetic-field conditions.[10] However, these high-order features were not thoroughly discussed. Understanding the contribution of these high-order features to the PHE and AMR would be essential for gaining deeper insights into the electronic transport properties and topological characteristics of t-PtBi$_2$. Therefore, we conducted a comprehensive study of the high-order features in the PHE and AMR of PtBi$_2$.

In this work, the magnetotransport properties of t-PtBi$_2$ single crystal are systematically measured. For the main feature in PHE and in-plane AMR curves, the magnetic field dependence of their amplitudes is close to $H^2$. Combined with the $\rho_{xx}$–$\rho_{xy}$ graph and the magnetoresistance behavior under different magnetic fields, we conclude that in addition to the chiral anomaly contributing to PHE, in-plane OMR also plays an important role, consistent with previous work.[10] Furthermore, both PHE and AMR show high-order features under low temperatures and strong magnetic fields, different from the magnetic field-induced changes in the AMR due to the topological phase transitions in other Weyl semimetals.[32,33] These features demonstrate similar temperature and magnetic field dependence as the turn-on behavior observed in resistivity–temperature $\rho(T)$ curves, indicating a shared physical origin for them. Additional measurements of MR at different angles, together with consistent theoretical calculations, indicate that these high-order features arise from the configuration of the Fermi surface, and the scaling behavior of MR renders them observable.

t-PtBi$_2$ single crystals were grown by the self-flux method. Platinum powder and bismuth powder were mixed thoroughly in a molar ratio of 1:8, placed in an alumina crucible, and then sealed in an evacuated quartz tube. The quartz tube was quickly heated to 800°C in a muffle furnace, and then left for 24 h, before being slowly cooled to 430 °C (2 °C/h). The excess amount of Bi was centrifuged at 430 °C. An integrated physical property measurement system (PPMS-14 T) was performed to measure the anisotropic transport properties in the plane.

The electronic structure calculations were performed by using density functional theory (DFT) within the generalized gradient approximation (GGA) of the Perdew–Burke–Ernzerhof (PBE) functional by the Vienna *Ab-initio* simulation package (VASP).[34,35] The cutoff energy is fixed to 500 eV, and all structures are relaxed until the energy and the forces converge to $10^{-7}$ eV and 0.0001 eV/Å, respectively. The Monkhorst–Pack k-point meshes used are 9 × 9 × 5 for sampling the Brillouin zone. The numerical calculations of AMR are performed by WannierTools.[36] For the generation of maximal localized Wannier functions (MLWFs), we choose the p orbitals of Bi and d orbitals of Pt as the initial projections. The outer window and frozen energy window are used for the disentanglement procedure. The tight-binding model is obtained using the Wannier90 code.[37]

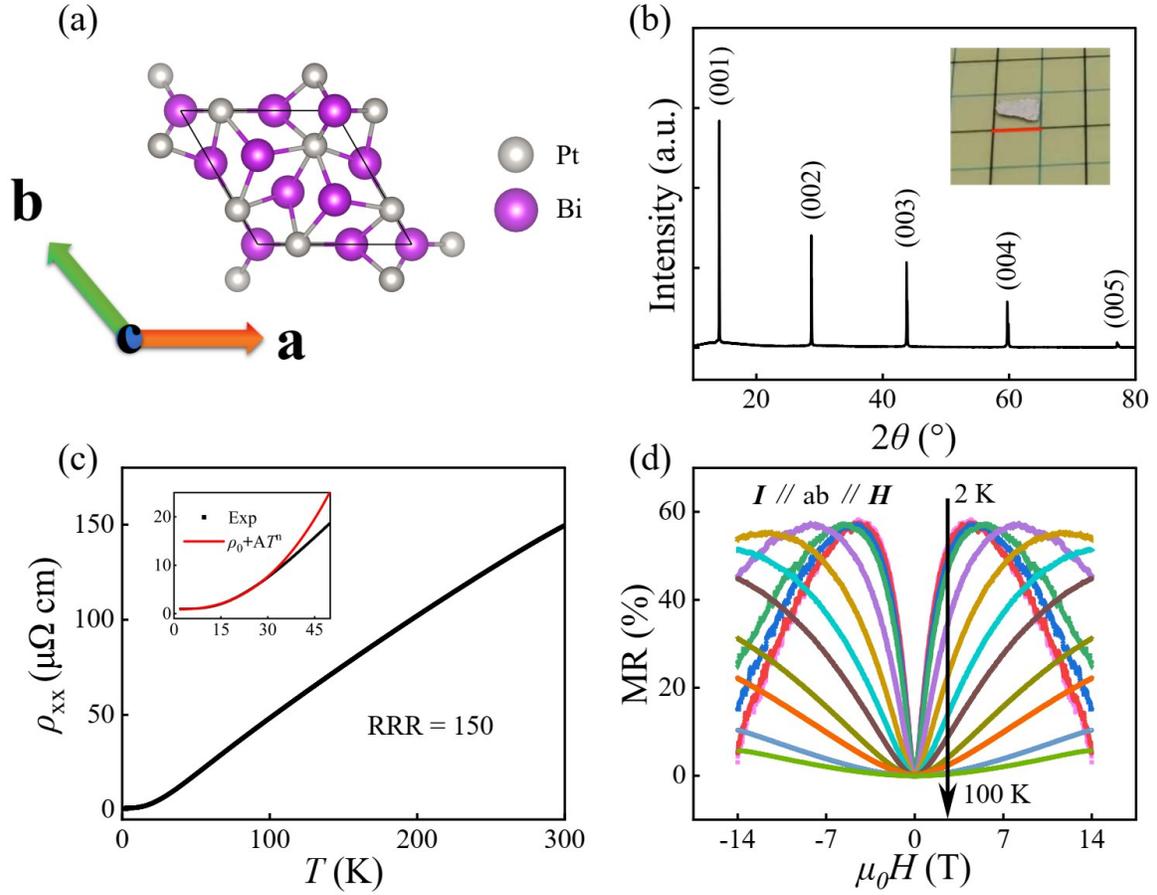

FIG. 1. (a) Crystal structure of t-PtBi$_2$. (b) Single crystal x-ray diffraction data on the surface of the sample (00$n$). The inset: optical micrograph of the t-PtBi$_2$ crystal, the scale bar is 5 mm. (c) Temperature dependence of longitudinal resistivity $\rho_{xx}$ of t-PtBi$_2$ samples with RRR = $R_{300K}/R_{2K}$ = 150. The inset: zoom-in of $\rho_{xx}(T)$ curve below $T$ < 50 K. (d) The variation of NMR with temperature when the magnetic field is parallel to the current ($I$//ab//$H$), from low to high temperatures of 2, 8, 10, 15, 20, 25, 30, 40, 50, 75, and 100 K.

The t-PtBi$_2$ is a layered material with a space group of $P31m$. Single-layer t-PtBi$_2$ is stacked along the direction of the c-axis and has rotational symmetry, as shown in Fig. 1(a). Figure. 1(b) shows the x-ray diffraction pattern of a single crystal of t-PtBi$_2$, where clear (00$n$) diffraction peaks can be observed, and no impurity peaks appear, indicating the high quality of the crystal. The metallic transport behavior of t-PtBi$_2$ is shown in Fig. 1(c), and

5 / 18

the residual resistivity [RRR = ($R_{300K}/R_{2K}$)] is as high as 150 with an absence of the magnetic field. In the high-temperature region, the approximately linear temperature dependence of $\rho_{xx}(T)$ indicates that electron–phonon (e–ph) scattering is the dominant scattering mechanism. Conversely, in the low-temperature region [as shown in the inset in Fig. 1(c)], it can be well fitted by using the formula $\rho_{xx}(T) = \rho_0 + AT^n$ with $\rho_0 = 0.89 \pm 0.01$ μΩ cm, A = $(1.36 \pm 0.08) \times 10^{-3}$ μΩ cm K$^{-2.5\pm0.02}$, and $n = 2.50 \pm 0.02$ are fitted quite well, which means that the resistance at low temperature arises from both the electron–electron scattering and electron–phonon scattering.[38] When the magnetic field is parallel to the current in the ab-plane, the NMR phenomenon appears [Fig. 1(d)], which is widely observed in topological materials due to the chiral anomaly.[39,40]

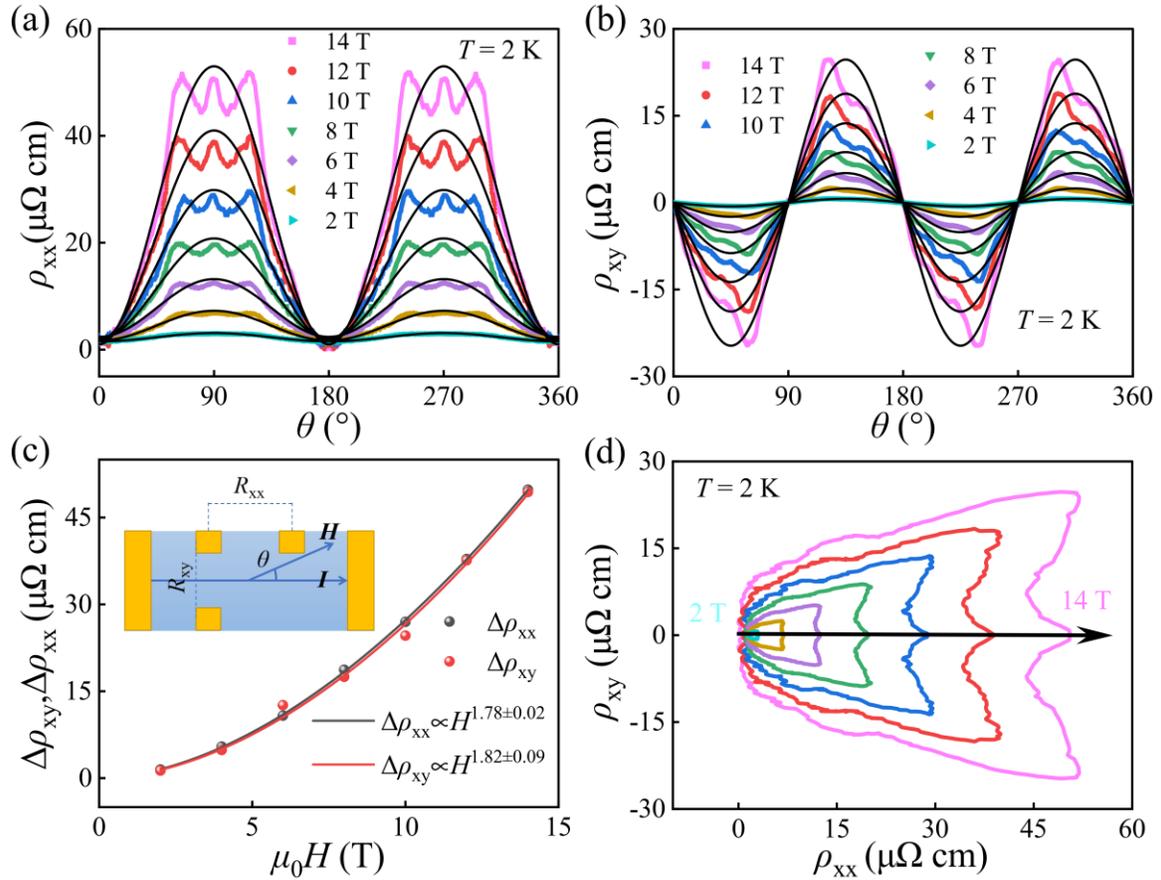

FIG. 2. (a) and (b) are AMR and PHE data (dots) and their fitting curves (solid lines) under different magnetic fields of 2–14 T, respectively. (c) The extracted $\Delta\rho_{xx}$ and $\Delta\rho_{xy}$ change (dots) with the magnetic

field. The red and black solid lines are the fitting lines of the $\Delta\rho_{xy}$ and $\Delta\rho_{xx}$, respectively. The inset of (c) shows a schematic diagram of the rotating magnetic field. (d) $\rho_{xy}$–$\rho_{xx}$ graphs under different magnetic fields shown in (a) and (b). The orbits evolve into a "shock-wave" pattern.

After completing the characterization of the sample's properties, we conduct systematic measurements of PHE and AMR on the sample with the magnetic field applied in the ab-plane, as shown in the inset of Fig. 2(c). Figures. 2(a) and 2(b) show the AMR and PHE of t-PtBi$_2$ under different magnetic fields at $T = 2$ K, where $\theta$ is the angle between the magnetic field and the current in the ab-plane. The resistivity tensor can be mainly captured by semi-classical Boltzmann theory:[19]

$$\rho_{xx} = \rho_\perp - \Delta\rho \cos^2\theta \tag{1}$$

$$\rho_{xy} = -\Delta\rho \sin\theta \cos\theta \tag{2}$$

where $\Delta\rho = \rho_\perp - \rho_\parallel$, with $\rho_\perp$ and $\rho_\parallel$ represent the resistivity of the magnetic field and current perpendicular (90°) and parallel (0°), respectively. $\rho_{xx}$ and $\rho_{xy}$ represent AMR and PHE when the magnetic field rotates in the plane, respectively. Both of them have a period of 180°, which is different from the ordinary Hall effect with a period of 360°. By fitting the main peaks, the field dependence exponent of $\Delta\rho_{xx}$ and $\Delta\rho_{xy}$ can be obtained as $1.78 \pm 0.02$ and $1.82 \pm 0.09$, respectively, as illustrated in Fig. 2(c). Theoretical calculations suggest that if PHE is solely attributed to the chiral anomaly, $\Delta\rho_{xy}$ would scale proportionally with $H^2$.[19] This slight exponential deviation indicates that, in addition to the chiral anomaly effect, there are other physical mechanisms contributing to PHE, such as the partial contribution of OMR in PHE according to the previous reports.[10] To gain a further understanding of the origin of PHE in t-PtBi$_2$, we plotted orbit parametric graphs of $\rho_{xx}$ and $\rho_{xy}$ against the angle $\theta$ at specific magnetic fields, as depicted in Fig. 2(d). All curves are enclosed by the curve for 14 T, presenting a pattern of quasi-concentric circles. However, the orbit parametric graphs expand rightward as the magnetic field increases, evolving into a shock-wave pattern.[41] This result further confirms that the PHE of t-PtBi$_2$ results from a combined effect of both the chiral anomaly and in-plane OMR.

It is worth noting that in the measurement results of AMR and PHE, there are some high-order features, and extra peaks around 60° and 120°, which cannot be well fitted by the aforementioned formula. These high-order features become more prominent as the magnetic field increases, and almost disappear at a low magnetic field, showing obvious magnetic field dependence. To explore the temperature dependence of the high-order features in AMR and PHE, their characteristic curves at different temperatures were measured when $\mu_0 H$ = 14 T. Figures. 3(a) and 3(b) show the behaviors of AMR and PHE from 2 to 100 K, where the amplitudes of $\rho_{xx}$ and $\rho_{xy}$ decrease dramatically with the increase in temperature. The high-order features also gradually weaken and completely undetectable at about 30 K.

We also measured the temperature dependence of $\rho_{xx}$ under different in-plane magnetic fields perpendicular to the current, as shown in Fig. 3(c). For $\mu_0 H$ = 14 T, the resistivity of the sample decreases with decreasing temperature in the high-temperature zone, achieving a minimum at approximately 30 K. Upon further cooling, the resistivity progressively rises, ultimately leveling off to saturation in the temperature range below 10 K, manifesting as a resistivity plateau. A similar turn-on behavior occurs as the magnetic field is parallel to the c-axis in t-PtBi$_2$,[2] which is widely observed in other semimetals.[42–46] To accurately determine the minimum value ($T_m$) of $\rho(T)$, we differentiated $\rho(T)$ and identified the temperature value at which $d\rho/dT = 0$ as $T_m$. As shown in the inset in Fig. 3(c), the $T_m \sim$ 28 K for $\mu_0 H$ = 14 T, shares a similar temperature at which the high-order features disappear in AMR and PHE[Figs. 3(a) and 3(b)], implying that they may have the same origin. We also measured the temperature-dependent characteristic curves of PHE at $\mu_0 H$ = 6 T. As shown in the supplementary material S1, the temperature where the high-order features of PHE and AMR disappear is about 20 K, which is also basically close to $T_m \sim 18$ K for 6 T obtained from Fig. 3(c). Hence, we deduce that the resistivity turn-on phenomenon and the high-order contributions in AMR and PHE share a common underlying physical mechanism, which is influenced by both temperature and magnetic field.

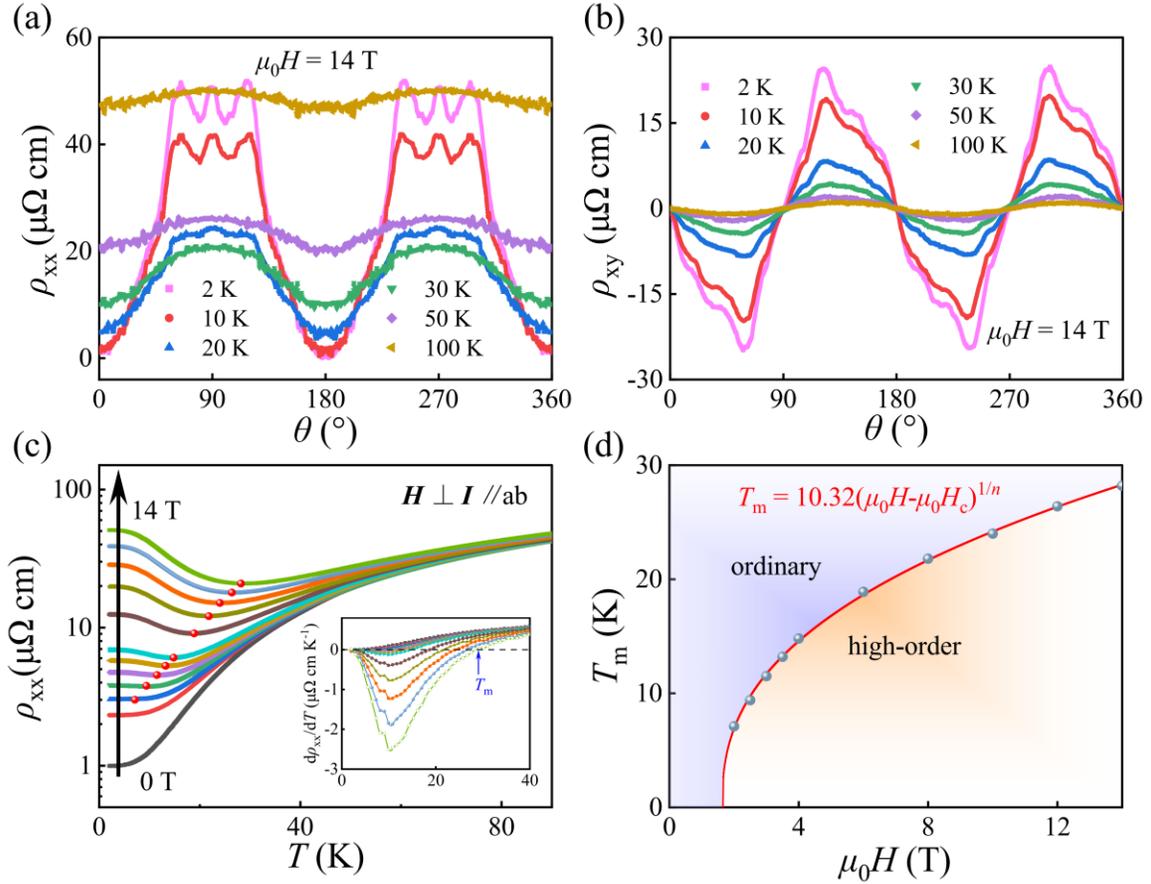

FIG. 3. The AMR (a) and PHE (b) at different temperatures under $\mu_0 H = 14$ T, respectively. (c) Temperature dependence of resistivity $\rho_{xx}$ measured under different magnetic fields with the in-plane magnetic field perpendicular to current. $\mu_0 H = 0, 1.5, 2, 2.5, 3, 3.5, 4, 6, 8, 10, 12,$ and 14 T. Inset: the plot of the derivative of resistivity to temperatures. The temperature corresponding to $d\rho/dT=0$ is defined as $T_m$. (d) A phase diagram based on the temperature-field dependence of the resistance minima in (c).

We also analyzed the $\rho(T)$ curves at other magnetic fields and observed a decrease in the corresponding $T_m$ values as the magnetic field weakened. Specifically, when the magnetic field diminished to 1.5 T, the turn-on behavior of resistivity in the low-temperature range ceased to exist entirely. This leads us to conclude that the turn-on phenomenon of resistivity is influenced by both temperature and magnetic field and so are the high-order features in AMR and PHE: in the presence of a fixed magnetic field $\mu_0 H_0$, the high-order features in AMR and PHE will emerge solely when the conditions $T < T_m$ and $\mu_0 H > \mu_0 H_0$ are met; otherwise, they remain absent. Based on the dependence of $T_m$ on the magnetic field shown in Fig. 3(c), a phase diagram is drawn in Fig. 3(d). By fitting the curve of $T_m$ and $\mu_0 H$ based on the formula $T_m = k(\mu_0 H - \mu_0 H_c)^{1/n}$, here $k = 10.32 \pm 0.23$ K/(T)$^{1/n}$, $n = 2.49 \pm 0.06$, consistent with the temperature exponent for the zero-field resistivity dependence $\rho_{xx}(T) = \rho_0 + AT^{2.50 \pm 0.02}$ shown in Fig. 1(c). According to the fitting results, we can empirically obtain a critical magnetic field $\mu_0 H_c \approx 1.63 \pm 0.05$ T. This implies that when the magnetic field is below $1.63 \pm 0.05$ T, the resistivity will not exhibit the turn-on phenomenon in the low-temperature region at any temperature, consistent with the absence of $T_m$ in the $\rho$–$T$ curve at 0 T and 1.5 T. We can further deduce that, for $\mu_0 H < 1.63 \pm 0.05$ T, notable high-order features will not be observed in both AMR and PHE even at $T = 0$ K. To confirm this deduction, we performed experiments on AMR and PHE at a temperature of 2 K with an external magnetic field of 1.5 T, and no clear high-order features were observed, as shown in Fig. S1, reinforcing the credibility of the phase diagram we have obtained.

In earlier scholarly works, the turn-on phenomenon observed in $\rho(T)$ and the high-order features of AMR (PHE) were documented as two distinct experimental phenomena. Typically, the emergence of the turn-on in $\rho(T)$ at low temperatures in the presence of a magnetic field, can be attributed to two physical mechanisms: one being the occurrence of a metal–insulator transition,[42,45] and the other being the scaling behavior of magnetoresistance, which is governed by a power-law dependence on both magnetic field and temperature.[47,48] Here, we tend to explain the turn-on in $\rho(T)$ using the second physical mechanism. First, t-PtBi$_2$, as a type of Weyl semimetal, has not been reported to undergo a metal-insulator transition at low temperatures thus far. Second, as shown in Fig. S2, the MR at 2 K is approximately proportional to $H^{1.7}$, where this exponent parameter is greater than 1, satisfying the criteria for the emergence of the turn-on in $\rho(T)$ under a magnetic field.[48]

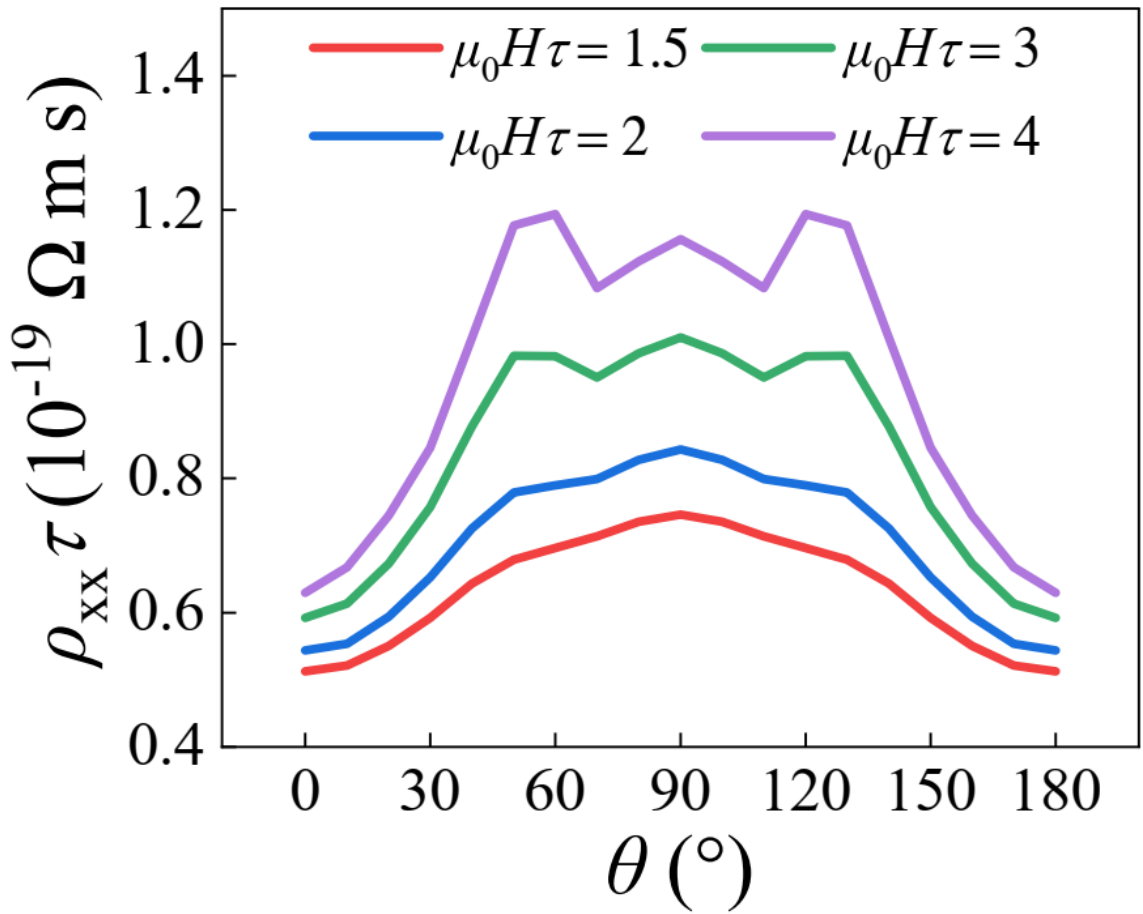

FIG. 4. The calculated AMR of the t-PtBi$_2$ single crystal at different $\mu_0 H \tau$ values. Here, the magnetic field is rotating in the ab-plane.

We also calculated the AMR of t-PtBi$_2$ based on the Fermi surface derived from DFT calculations in conjunction with Boltzmann transport theory.[49] Figure 4 displays the calculated resistivity $\rho_{xx}$ at different $\mu_0 H\tau$ values, with the electric field aligned along the a-axis and the magnetic field rotating within the ab-plane. Here, $\tau$ is the relaxation time of the band. In this figure, 0° represents the alignment where the magnetic field is parallel to the electric field. Notably, when we fix the temperature, it implies that $\tau$ is a constant. Then, high-order features emerging at 60° and 120° intensify as the magnetic field increases, following a trend similar to that observed in Fig. 2(a). Upon fixing the magnetic field and elevating the temperature, the $\tau$ value decreases with increasing temperature. Consequently, the high-order features of AMR progressively become unrecognizable as the temperature rises. Our experimental observations shown in Figs. 2 and 3 align with these theoretical results in terms of the dependence of the high-order features of AMR on both temperature and magnetic field, demonstrating that the high-order features originate from the intrinsic band structure of t-PtBi$_2$. Previous literature has also suggested that the presence of high-order features in AMR (PHE) is related to the configuration of the Fermi surface,[50] which we have confirmed through the analysis of Kohler's rule applied to MR in different directions for t-PtBi$_2$ (Fig. S2). The anisotropic Fermi surface of t-PtBi$_2$ can be found in the supplementary materials (Fig. S3), which is consistent with the results in previous literature.[2,51] Considering the phase diagram depicted in Fig. 3(d), we can further assert that the turn-on phenomenon in the $\rho(T)$ curve arises from the material's intrinsic properties in response to temperature and magnetic field variations. Furthermore, our findings indicate that the observations of high-order features in AMR (PHE) are influenced not only by the anisotropy of the Fermi surface but also by the scaling behavior of MR in t-PtBi$_2$.

In summary, we have observed distinct high-order features of the PHE and AMR in t-PtBi$_2$ bulk samples under low-temperature and high-magnetic-field conditions. These high-order features are closely related to the turn-on phenomenon observed in $\rho(T)$ curves at a fixed magnetic field. Based on our observations, we have constructed a magnetic field and temperature phase diagram to illustrate the observation of high-order features in AMR and PHE. The fitting results indicate that when the magnetic field is below 1.63 ± 0.05 T, the $\rho(T)$ curve will not exhibit resistance turn-on in the low-temperature region, and neither PHE nor AMR will display clear high-order features. Furthermore, our theoretical calculations of AMR show a trend consistent with experimental results, where high-order features are significant at high fields and low temperatures. This demonstrates that the detection of high-order features of PHE and AMR in t-PtBi$_2$ can be attributed to the combined contributions of Fermi surface anisotropy and the scaling behavior of MR. Our work will provide deeper insights into the origin of high-order PHE in the non-magnetic topological semimetal. The significant and complex PHE features make t-PtBi$_2$ promising to be a good platform for designing plane Hall sensors, especially 3D compacted "Lab-on-a-chip" devices.[52]

See the supplementary material for additional details related to this work.

This work was supported by the National Natural Science Foundation of China (Grants Nos. 12274278, 12204300, 52130204, 12311530675, and 12074241), the Capacity Building for Sci-Tech Innovation-Fundamental Scientific Research Funds [20530290057 (H.W.)], the Science and Technology Project of Beijing Municipal Education Commission [KM202010028014 (H.W.)], Science and Technology Commission of Shanghai Municipality (No.22XD1400900), and High Performance Computing Center, Shanghai University.

# AUTHOR DECLARATIONS

## Conflict of Interest

The authors have no conflicts to disclose.

## Author Contributions

Fangqi Cai and Mingxi Chi contributed equally to this work.

**Fangqi Cai**: Formal analysis (equal); Investigation (equal); Methodology (equal); Writing – original draft (equal). **Mingxi Chi**: Investigation (equal); Methodology (equal). **Yingjie Hu**: Formal analysis (equal). **Heyao Liu**: Methodology (equal). **Yangyang Chen**: Conceptualization (equal). **Chao Jing**: Funding Acquisition (equal). **Wei Ren**: Formal analysis (equal); Funding Acquisition (equal); Methodology (equal). **He Wang**: Conceptualization (equal); Project administration (equal); Supervision (equal); Writing/Review & Editing (equal).

# DATA AVAILABILITY

The data that support the findings of this study are available from the corresponding authors upon reasonable request.